# Waveguiding in two-dimensional Floquet non-Abelian topological insulators

Yujie Zhou,[1] Changsen Li,[1] Xiumei Wang,[2*] and Xingping Zhou [3*]

[1] *College of Integrated Circuit Science and Engineering, Nanjing University of Posts and Telecommunications, Nanjing 210003, China*

[2] *College of Electronic and Optical Engineering, Nanjing University of Posts and Telecommunications, Nanjing 210003, China*

[3] *Institute of Quantum Information and Technology, Nanjing University of Posts and Telecommunications, Nanjing 210003, China*

[*]*Author to whom any correspondence should be addressed.*

[*]*wxm@njupt.edu.cn*
[*]*zxp@njupt.edu.cn*

Topological phases characterized by non-Abelian charges have garnered increasing attention recently. Although Floquet (periodic-driving) higher-order topological phases have been explored at the single-particle level, the role of interactions in non-Abelian topological insulators with multiple entangled energy gaps remains incompletely understood. In this work, we extend previous research by investigating higher-order topological phases featuring non-Abelian charges through Floquet engineering. Here we construct a model for two-dimensional non-Abelian higher-order topological phases on a square lattice subjected to two-step periodic driving. We find that the corner and edge states emerge and appear in all energy gaps despite the quaternion charge being trivial ( $q=1$ ). Moreover, spatially exchanging the driving generates exotic interface modes—a hallmark of non-Abelian dynamics, namely non-commutativity. Notably, the non-zero composite Chern number demonstrates the non-triviality of the Floquet non-Abelian system with $q=1$. We further reveal that the configuration of these quaternion-charge edge states is entirely determined by the quadruple degenerate phase-band singularities in the time evolution. Our work provides a platform for studying higher-order topological states and non-equilibrium quantum dynamics.

# I. INTRODUCTION

Topological insulators are fascinating phases of matter that exhibit insulating behavior in their $d$-dimensional bulk, yet support conduction along their $(d-1)$-dimensional boundaries, where gapless boundary states emerge within the gapped bulk energy spectrum [1-5]. The boundary states are determined by the topological properties of the bulk bands—a principle known as the bulk–boundary correspondence (BBC). The introduce of non-Hermiticity profoundly modifies the topological properties, leading to unprecedented phenomena beyond the descriptions of Bloch band theory, e.g., the breakdown of the conventional bulk-boundary correspondence. The non-Bloch band theory [6-10], biorthogonal bulk-boundary correspondence [11, 12], and the non-Bloch Band theory in arbitrary dimensions [13, 14] are proposed to solve this problem.

Another intriguing avenue in the pursuit of exotic phases is Floquet engineering, in which a system is periodically driven and probed via stroboscopic measurements. Novel topological phenomena without static counterparts have been predicted, such as the anomalous Floquet topological (AFT) phase [15-18]. In this phase, all quasienergy bands exhibit zero Chern numbers, yet robust edge modes persist within the gaps. A modified bulk-boundary correspondence (BBC) is constructed using winding numbers defined in time–momentum space to characterize the topological properties of quasienergy gaps, where these winding numbers are gap-dependent and distinct from one another [18-20]. The topological Floquet states interference [21-25], anomalous skin modes [26-31], and higher-order Floquet topological phases [32-35] have been observed in various periodic driving systems. Besides, high-frequency drives can induce band inversion and produce topological gaps at Dirac points [36-38].

Very recently, it has been found that Floquet driving may endow multigap systems with non-Abelian topology, which is called Floquet non-Abelian topological insulator (FNATI) [15-17, 39-44]. The interface effect resulting from swapped driving has been uncovered in FNATI and the BBC topological properties can be characterized by phase-band singularities. In Floquet non-Abelian topological systems, when band singularities and braiding phenomena are considered in a gap-dependent manner, topological invariants distinct from Chern numbers are generated [42-47], transcending the conventional topological classification based on single energy gaps. As we know, the advent of higher order topological insulators (HOTIs) [48-52] has attracted enormous interest in the modern condensed-matter physics community. The realization of Floquet HOTIs has also been investigated from different perspectives in recent literature [25, 26, 53]. For Floquet engineering, most studies treat the periodicity separately from the spatial one. In fact, the periodic-driving system combined with spatial varying can exhibit much richer phenomenon than a simple Floquet product [54-59].

In this work, we construct a model for two-dimensional non-Abelian higher-order topological phases on a square lattice subjected to two-step periodic driving. The static and Floquet driving conditions are both considered in the presence of spatial dislocations. We employ the gap-dependent composite Chern number to characterize the topological properties of non-Abelian systems with degenerate energy bands. During time evolution, the opening and closing of phase bands induce the formation of Dirac points, which in turn determine the presence or absence of boundary states within the bandgap.

## II. MULTI-GAP TOPOLOGY AND DRIVING PROTOCOL

### A. Static three-band topological insulator with quaternion charge $Q_8$

In the presence of parity-time (PT) symmetry, when expressed in an appropriate basis, the Hamiltonian becomes real-valued in momentum space, $H(k) = H^*(k)$. The $Q_8$ [60] contains 8 elements and 5 conjugacy classes $\{1, \pm i, \pm j, \pm k, -1\}$. In our work we construct the following structure based on $q = 1$. We consider a two-dimensional lattice in Fig. 1(a), where each unit cell contains three lattice sites denoted as A, B, and C. The Floquet driving is based on two components, namely $H_1$ and $H_2$ as shown in Fig. 1(b). The $H_1$ denotes that the unit cells in the $x \times y$ 2D structure only undergo intra-cell coupling individually, while the $H_2$ denotes that the unit cells in the $x \times y$ 2D structure undergo inter-cell coupling sequentially in a certain order.

Without loss of generality, the driving period is set to $T = 1$. We employ a symmetric driving protocol and do not consider the spatial exchange driving, as shown in Fig. 1(a):

$$H = H_1 = \sum_{i,j} \sum_{M,N} s_{MN} c^\dagger_{M,i,j} c_{N,i,j} + \text{h. c.} \tag{1}$$

for $t \in mT + [0, T/4] \cup [3T/4, T]$ $(m \in \mathbb{Z})$ and

$$H = H_2 = \sum_{i=1}^{x-1}\sum_{j=1}^{y}\sum_{M,N} v_{MN} c^{\dagger}_{M,i,j} c_{N,i+1,j} + \text{h.c.}$$
$$+ \sum_{i=1}^{x}\sum_{j=1}^{y-1}\sum_{M,N} v_{MN} c^{\dagger}_{M,i,j} c_{N,i,j+1} + \text{h.c.} \qquad (2)$$

for $t \in mT + [T/4, 3T/4]$ ($m \in \mathbb{Z}$). Here, $c^{\dagger}_{M,n}$ and $c_{M,n}$ denote the creation and annihilation operators at the lattice site M (M=A, B, C) and N (N=A, B, C) in the $n$-th unit cell, respectively. The $x, y \in \mathbb{Z}$ represent the number of lattice points corresponding to each row and each column respectively. The coupling parameters $s_{MN}$ and $v_{MN}$ used in this work are as follows: $s_{AA} = s_{BB} = s_{CC} = 0$, $s_{AB} = s_{BA} = r$, $s_{BC} = s_{CB} = s$, $s_{CA} = s_{AC} = t$, $v_{AB} = v_{BA} = iu$, $v_{BC} = v_{CB} = iv$, and $v_{CA} = v_{AC} = iw$. Since these coupling parameters correspond to a static three-band topological insulator characterized by the non-Abelian charge $q = 1$, we have $r = 1$, $s = 0$, $t = 3$, $u = 0$, $v = 0$, $w = -3$, $v_{AA} = -2$, $v_{BB} = 0$, and $v_{CC} = 2$.

The dynamics of the system are governed by the time evolution operator $U(t) = \text{T} \exp\left(-i\int_0^t H(\tau)d\tau\right)$, where $\text{T}$ denotes time ordering. The stroboscopic evolution of the system is described by the Floquet operator,

$$U(T) = e^{-iH_1 T/4} e^{-iH_2 T/2} e^{-iH_1 T/4} \qquad (3)$$

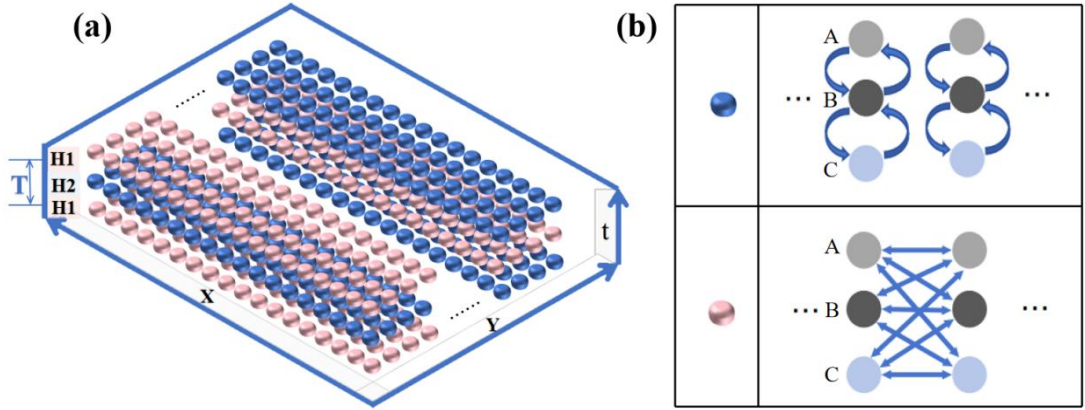

**FIG. 1.** Models of two-dimensional PT-symmetric Floquet non-Abelian topological lattices. (a) The model under a symmetric driving protocol. A lattice with size of $x \times y$ undergoes $H_1$ and $H_2$ exchange driving along the time axis(t). (b) Building blocks for our Floquet driving. The blue ball contains only intracell coupling terms, the pink ball contains only intercell coupling terms.

The effective Floquet Hamiltonian $H_F$ is defined via the relation $U(T) = e^{-iH_F T}$, where $U(T)$ denotes the time evolution operator over one driving period $T$. Owing to the symmetric nature of the driving protocol, the Floquet Hamiltonian exhibits PT symmetry, which is mathematically expressed as $H_F(k) = H_F^*(k)$. Upon diagonalization of $H_F$, i.e., $H_F |u_n\rangle = \varepsilon_n |u_n\rangle$ (where n serves as the band index), we obtain the quasienergies $\varepsilon_n$ and their corresponding eigenstates $u_n$. It is noteworthy that the quasienergies are well-defined only modulo $2\pi/T$, giving rise to the formation of quasienergy bands. In the subsequent analysis, we restrict the quasienergies to the first Brillouin zone (FBZ) of quasienergy, such that $\varepsilon_n \in (-\pi/T, \pi/T]$.

### B. BBC and interface modes induced by swapped driving

To investigate the topological properties of 2D Floquet non-Abelian systems, we separately vary the temporal driving protocol and spatial driving distribution of the 2D structure. The distinct evolutionary behaviors of unit cells in space and time can be understood through the simplified classification diagram shown in Fig. 2. The second and third rows represent the spatial distribution states under static and Floquet driving, respectively. Additionally, the third row briefly illustrates the temporal evolution of different structures with schematic diagrams. We next primarily focus on the significance of the third row in the schematic diagram. The second and third columns describe the evolution sequences of the global Hamiltonian as $H_1 \to H_2 \to H_1$ and $H_2 \to H_1 \to H_2$, respectively. We denote these two bulk Floquet operators as $U_1$ and $U_2$, which are related via a similarity transformation: $U_1 = V^{-1} U_2 V$, where $V = e^{iH_2 T/4} e^{iH_1 T/4}$ accounts for the time shift. This shift does not alter the quasienergy spectrum regardless of the boundary conditions. For the fourth and seventh columns, we consider a system with swapped driving sequences on the two sides. Within one complete period, the Hamiltonian evolution sequences on the left (upper) side and the right (lower) side are $H_1 \to H_2 \to H_1$ and $H_2 \to H_1 \to H_2$, respectively. The bulk Floquet operators on the two sides also satisfy $U_1$ and $U_2$. For the fifth and sixth columns, we consider a system with a unilaterally swapped driving sequence. Specifically, in the fifth column, the Hamiltonian evolution sequences on the left and right sides are $H_1 \to vacuum \to H_1$ and $vacuum \to H_1 \to vacuum$, respectively. In the sixth column, the Hamiltonian evolution sequences on the right and left sides are $H_2 \to vacuum \to H_2$ and $vacuum \to H_2 \to vacuum$, respectively. The Hamiltonian driving scenarios in the eighth and ninth columns are similar to those in the fifth and sixth columns, except that the left-right distribution is replaced by an upper-lower distribution. When the spatial distribution is vacuum, we represent it with the zero matrix $H_{vacuum} = \mathbf{0}$.

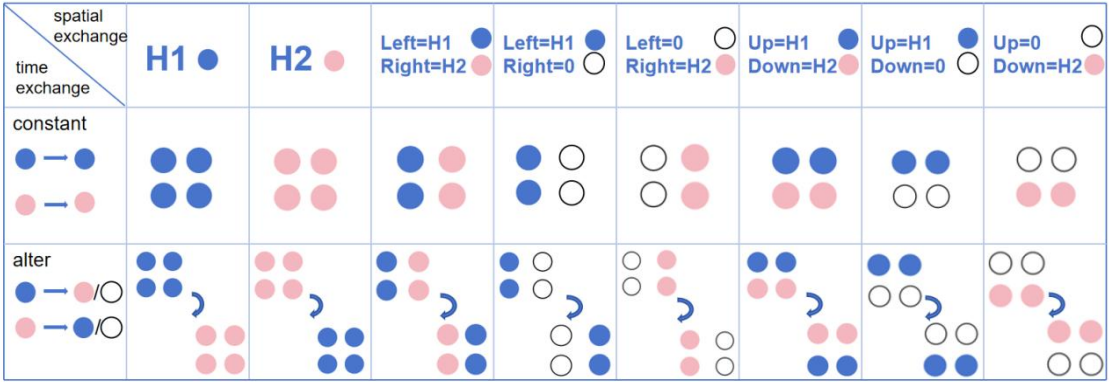

**FIG. 2.** Diverse spatial and temporal distributions of quaternion charge $q = 1$. The first row presents the unit cell structures corresponding to distinct spatial distributions in the 2D architecture, namely $H_1$, $H_2$, and the zero matrix. The second row describes the 2D spatial distribution of unit cells under static conditions. The third row describes the dynamic evolution of the 2D spatial distributions of different unit cells under Floquet time driving.

We systematically vary only one variable at a time to better investigate the edge states, corner states, and interface modes induced by various modes as depicted in Fig. 2. The variable can be either time exchange or spatial exchange. The periodic evolution of the unit cells can be understood through the simple heuristic analysis in Fig. 3. The unit cells labeled in bold font represent the boundary state portion relative to the original structure. Each schematic diagram corresponds to its nearest neighboring driven system and the associated special eigenstates. All primitive cells are labeled as B (bulk), E1 (edge-1), E2 (edge-2), and C (corner) according to their positions. We describe the differences in the Hamiltonian evolution of each system through different combinations of time exchange and spatial exchange.

As shown in Fig. 3(a), we describe the static mode [61] where the Hamiltonian is $H_2$, and $x = y = 18$. In this case, the stroboscopic evolution of the system can be described by $U_S(T) = e^{-iH_2 T}$. The quasienergies of this phase and the spatial profiles of their associated eigenstates are plotted, in which distinct corner states and boundary states are clearly observed. In this scenario, the boundary states manifest in two distinct forms: one with energy concentrated at edge-1 and the other with energy concentrated at edge-2. The simultaneous energy concentration at both edge-1 and edge-2 does not occur. In Fig. 3(a), we only show the case where energy is concentrated at edge-2. The energy localization in the eigenstate diagrams for edge-1 is presented in Appendix A of the Supplementary Material. A similar phenomenon also occurs when the Hamiltonian on one side is $H_2$ and that on the other side is a zero matrix. It is because that the zero matrix represents a vacuum, and in the case where only the Hamiltonian $H_2$ is present, its external environment can also be regarded as a vacuum. This scenario corresponds to the three cases illustrated in the 3rd, 6th, and 9th columns of Fig. 2. In these cases, when considering the corresponding Hamiltonian with only intra-cell coupling, energy (or quasienergy) undergoes sequential accumulation within each unit cell as it changes. The behavior occurs in both static and Floquet-driven systems, without the emergence of any noticeable corner states or edge states.

If a time exchange is applied to an originally static system, making the evolution sequences of its global Hamiltonian be $H_1 \to H_2 \to H_1$ or $H_2 \to H_1 \to H_2$, it is converted into the eigenstates corresponding to Fig. 3(b). At this case, $x = y = 18$. The stroboscopic evolution of the system satisfies:

$$\begin{cases} U(T) = e^{-iH_1 T/4} e^{-iH_2 T/2} e^{-iH_1 T/4}, \\ H_1 = \sum_{i,j} \sum_{M,N} s_{MN} c^\dagger_{M,i,j} c_{N,i,j} + \text{h. c.}, \\ H_2 = \sum_{i=1}^{17} \sum_{j=1}^{18} \sum_{M,N} v_{MN} c^\dagger_{M,i,j} c_{N,i+1,j} + \text{h. c.} + \sum_{i=1}^{18} \sum_{j=1}^{17} \sum_{M,N} v_{MN} c^\dagger_{M,i,j} c_{N,i,j+1} + \text{h. c.} \end{cases} \quad (4)$$

We can observe that its boundary states and corner states, emerge in pairs and exhibit symmetry, unlike those in static systems.

We incorporate spatial exchange on the basis of Hamiltonian time exchange in Fig. 3(c). Within one complete period, the Hamiltonian evolution sequences on left and right side are $H_1 \to H_2 \to H_1$ and $H_2 \to H_1 \to H_2$, respectively. It should be noted that $x = 18$ and $y = 9$ at this point. The stroboscopic evolution of the

system also satisfies:

$$\begin{cases} U_1(T) = e^{-iH_1T/4}e^{-iH_2T/2}e^{-iH_1T/4}, \\ U_2(T) = e^{-iH_2T/4}e^{-iH_1T/2}e^{-iH_2T/4}, \\ H_1 = \sum_{i,j}\sum_{M,N} s_{MN} c^{\dagger}_{M,i,j} c_{N,i,j} + \text{h. c.}, \\ H_2 = \sum_{i=1}^{17}\sum_{j=1}^{9}\sum_{M,N} v_{MN} c^{\dagger}_{M,i,j} c_{N,i+1,j} + \text{h. c.} + \sum_{i=1}^{18}\sum_{j=1}^{8}\sum_{M,N} v_{MN} c^{\dagger}_{M,i,j} c_{N,i,j+1} + \text{h. c.} \end{cases} \quad (5)$$

$U_1(T)$ and $U_2(T)$ represent the stroboscopic evolution of the left and right subsystems, respectively.

Under the bulk state correspondence guaranteed by PT symmetry, the $SO(3)$ transformation $O$ corresponds to the non-identity element in its fundamental group $\pi_1(SO(3)) = \mathbb{Z}_2$ [62]. In this case, the quaternion charges on both sides satisfy $q_L = -q_R$, which implies $q_L \neq q_R$ [63]. A topological mismatch emerges within the bandgaps when a non-eliminable (by symmetric operations) difference arises in the non-Abelian topological charges (i.e., quaternion charges $q_L \neq q_R$) between two non-Abelian subsystems. This mismatch induces stable interface modes at the interface. Thus, under Floquet driving with spatial exchange, not only boundary states and corner states emerge, but also novel interface modes appear. Interface modes arise as a consequence of the distinct driving sequences of the left and right subsystems in non-Abelian systems, which reflects the noncommutativity inherent [64] in such systems. In contrast, interface modes do not emerge in Abelian systems [65], since the driving sequence has no impact on the outcomes in these systems. These special eigenstates can occur along edge-1 or edge-2, owing to the spatially exchanged driving. Here, we only display the special eigenstates with energy concentrated on the right side. The eigenstates with energy concentrated on the left side are shown in Appendix A of the Supplementary Material. The phenomenon in Fig. 3(d) also aligns with such a scenario, wherein the spatial exchange distribution transitions from a left-right configuration to an upper-lower configuration. At this point, $x = 9$ and $y = 18$. The stroboscopic evolution of the system satisfies:

$$\begin{cases} U_1(T) = e^{-iH_1T/4}e^{-iH_2T/2}e^{-iH_1T/4}, \\ U_2(T) = e^{-iH_2T/4}e^{-iH_1T/2}e^{-iH_2T/4}, \\ H_1 = \sum_{i,j}\sum_{M,N} s_{MN} c^\dagger_{M,i,j} c_{N,i,j} + \text{h. c.}, \\ H_2 = \sum_{i=1}^{8}\sum_{j=1}^{18}\sum_{M,N} v_{MN} c^\dagger_{M,i,j} c_{N,i+1,j} + \text{h. c.} + \sum_{i=1}^{9}\sum_{j=1}^{17}\sum_{M,N} v_{MN} c^\dagger_{M,i,j} c_{N,i,j+1} + \text{h. c.} \end{cases} \quad (6)$$

$U_1(T)$ and $U_2(T)$ represent the stroboscopic evolution of the upper and lower subsystems, respectively. We only present the special eigenstates with energy concentrated on the upper side. The eigenstates with energy concentrated on the lower side can be found in Appendix A of the Supplementary Materials.

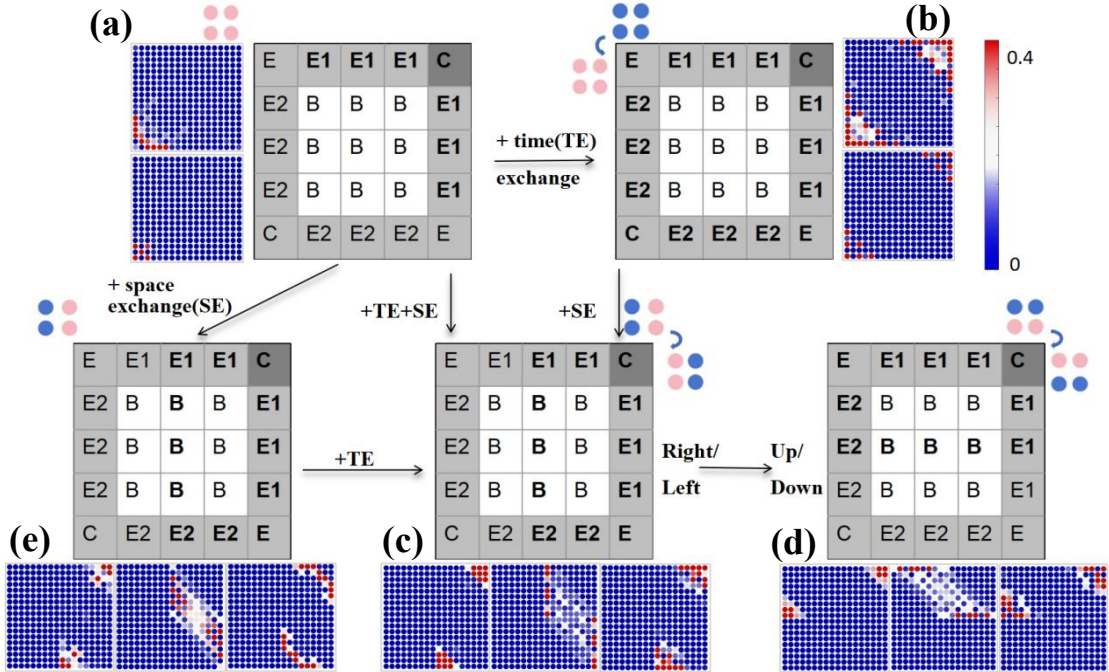

FIG. 3. BBC and interface modes induced by different driving modes. (a-b) Spatial distribution of special eigenstates in static and Floquet systems. The top-left panel depicts the edge states emerging during the evolution process, while the bottom-left panel illustrates the corner states. (c-e) Spatial distribution of special eigenstates in Floquet and static systems under spatial exchange driving. From left to right, they are the corner states, interface modes, and edge states emerging during the evolution, respectively.

If a spatial exchange operation is solely applied to an initially static system, resulting in left and right Hamiltonians (denoted as $H_1$ and $H_2$ respectively), it is revealed that interface modes also emerge in Fig. 3(e). Since the two subsystems have different coupling modes, they are regarded as two systems with distinct topological structures. This difference cannot be eliminated by symmetric operations. Meanwhile, the boundary states, corner states, and boundary modes emerging under such circumstances only appear on the side where the Hamiltonian is $H_2$.

## III. CHARACTERIZATION OF TOPOLOGICAL PHASES

### A. Composite Chern number

To investigate Floquet systems, we introduce classical topological invariants that can reveal the correspondence between the bulk topological properties of the system and its boundary states (such as edge states). Therefore, when studying Floquet systems, the first consideration is to introduce the Chern number to characterize their topological properties [66]. Given that the energy bands of systems with non-Abelian charges are degenerate, it is impossible to assign a Chern number to each individual energy band. In this scenario, these degenerate bands (taking the bands from $n$ to $n+N-1$ as an example) collectively possess a composite Chern number, denoted as $C(n \oplus n+1 \oplus \cdots \oplus n+N-1)$, which is linked to the multiplets $u_{nk} \oplus u_{n+1,k} \oplus \cdots \oplus u_{n+N-1,k}$ [67]. Next, we calculate the composite Chern number for the case corresponding to Fig. 3(b). The system is governed by a time-periodic Hamiltonian $H(t) = H(t+T)$ with period $T=1$. The stroboscopic evolution over one period is described by the Floquet evolution operator $U(\mathbf{k})$, where $\mathbf{k}=(k_x, k_y)$ denotes the 2D momentum. For a three-band tight-binding model with intra-cell static coupling $S$ and inter-cell drive-induced coupling $V(\mathbf{k})$, the evolution operator is constructed as:

$$U(\mathbf{k}) = \exp\left(-iS \cdot \frac{1}{4}\right) \cdot \exp\left(-iV(\mathbf{k}) \cdot \frac{1}{2}\right) \cdot \exp\left(-iS \cdot \frac{1}{4}\right) \qquad (7)$$

where $S$ is the 3×3 intra-cell coupling matrix:

$$S = \begin{bmatrix} s_{AA} & s_{AB} & s_{AC} \\ s_{BA} & s_{BB} & s_{BC} \\ s_{CA} & s_{CB} & s_{CC} \end{bmatrix} \qquad (8)$$

and $V(\mathbf{k})$ is the 3×3 momentum-dependent inter-cell coupling matrix:

$$V(k_x, k_y) = \begin{bmatrix} 2v_{AA}(\cos k_x + \cos k_y) & 2u(\sin k_x + \sin k_y) & 2w(\sin k_x + \sin k_y) \\ 2u(\sin k_x + \sin k_y) & 2v_{BB}(\cos k_x + \cos k_y) & 2v(\sin k_x + \sin k_y) \\ 2w(\sin k_x + \sin k_y) & 2v(\sin k_x + \sin k_y) & 2v_{CC}(\cos k_x + \cos k_y) \end{bmatrix} \qquad (9)$$

To compute the Wilson loop, we first solve the eigenvalue problem of $U(\mathbf{k})$:

$U(\mathbf{k})|u_n(\mathbf{k})\rangle = \lambda_n(\mathbf{k})|u_n(\mathbf{k})\rangle$ where $|u_n(\mathbf{k})\rangle$ ($n=1,2,3$) are the eigenstates, and $\lambda_n(\mathbf{k}) = \exp(-i\epsilon_n(\mathbf{k}))$ are the eigenvalues with quasienergies $\epsilon_n(\mathbf{k})$. Eigenstates exhibit gauge freedom (arbitrary phase factors), which must be fixed to ensure continuity in momentum space [33]. We adopt a phase gauge by fixing the phase of the first component of each eigenstate: $|u_n(\mathbf{k})\rangle \mapsto |u_n(\mathbf{k})\rangle \cdot \exp\left(-i\arg\left(u_n^{(1)}(\mathbf{k})\right)\right)$ where $u_n^{(1)}(\mathbf{k})$ is the first component of $|u_n(\mathbf{k})\rangle$, and $\arg(\cdot)$ denotes the phase angle. It ensures smooth variation of eigenstates across the FBZ. For multi-band systems, the Berry connection is generalized to a 3×3 matrix $\mathbb{A}(\mathbf{k})$, describing the geometric phase gradient of the full eigenstate manifold. For a discretized Brillouin zone (BZ), $\mathbb{A}(\mathbf{k})$ is approximated via finite differences [68]. For a given wavevector $\mathbf{k} = (k_x, k_y)$, the connection $\mathbb{A}(\mathbf{k})$ is evaluated along the $k_x$-direction. Specifically, at boundary points ($i=1$ or $i=N_1$): $\mathbb{A}(k_x^i, k_y^j) \approx -iU_i^\dagger \cdot \dfrac{U_{i\pm 1} - U_i}{\Delta k_x}$. For interior points ($2 \le i \le N_1 - 1$): $\mathbb{A}(k_x^i, k_y^j) \approx -iU_i^\dagger \cdot \dfrac{U_{i+1} - U_{i-1}}{2\Delta k_x}$. Here, $U_i = [|u_1(k_x^i, k_y^j)\rangle, |u_2(k_x^i, k_y^j)\rangle, |u_3(k_x^i, k_y^j)\rangle]$ represents the $3\times 3$ matrix of eigenstates at $k_x$, and $U_i^\dagger$ is its conjugate transpose. The parameter $N_1$ denotes the number of discrete points selected when discretizing the FBZ in the $k_x$-direction. And $\Delta k_x = 2\pi/N_1$ is the discretization step in the $k_x$-direction. The composite Wilson loop is defined as the matrix product of exponential phases accumulated along a closed path in momentum space. For each fixed $k_y^j$, we compute the Wilson loop along the $k_x$-direction (a closed loop covering $k_x \in [-\pi, \pi]$):

$$\mathbb{W}(k_y^j) = \prod_{i=1}^{N_1} \exp\left(\mathbb{A}(k_x^i, k_y^j) \cdot \Delta k_x\right) \tag{10}$$

This product is computed iteratively: starting with the identity matrix $\mathbb{W}_0 = I$, each step updates the loop via $\mathbb{W}_i = \exp\left(\mathbb{A}(k_x^i, k_y^j) \cdot \Delta k_x\right) \cdot \mathbb{W}_{i-1}$. The final result $\mathbb{W}(k_y^j) = \mathbb{W}_{N_1}$ is a 3×3 matrix encoding the total geometric phase accumulation around the $k_x$-loop at fixed $k_y^j$. The composite Chern number, a global topological invariant, is derived from the $k_y$-dependence of $\mathbb{W}(k_y^j)$. For adjacent points $k_y^j$

and $k_y^{j+1}$, we define the "link matrix" as:

$$L(k_y^j) = W(k_y^j)^{-1} \cdot W(k_y^{j+1}) \tag{11}$$

The determinant of $L(k_y^j)$ captures the combined phase change of all bands [69]. The composite Chern number is the total phase accumulated along the $k_y$-direction, normalized by $2\pi$:

$$C = \frac{1}{2\pi} \sum_{j=1}^{N_2-1} \arg\left(\det\left(L(k_y^j)\right)\right) \tag{12}$$

where $\arg(\cdot)$ denotes the phase of the determinant, and $N_2$ is the number of discretization points in $k_y$-direction. The derived composite Chern number quantifies the global topological order. Through the aforementioned rigorous theoretical calculations, we obtain the value of the quaternion charge topological invariant as $C = -1$. This non-zero value ($C \neq 0$) provides direct evidence for the topological non-triviality of the 2D non-Abelian Floquet system [70], indicating the existence of stable topologically protected edge states within the system.

**B. degenerate Dirac points in tangled bandgaps**

In this case, the composite Chern number can describe the overall topological properties of the structure but fails to capture the full picture of the edge states during time evolution. In fact, the emergence of edge states has a dynamical origin, and the complete information of dynamical topology is encoded in the full-time evolution operator. By introducing phase bands and their associated momentum-time singularities [33], a unified analysis of the BBC can be conducted. It should be noted that the time evolution operator $U(t)$ does not always satisfy PT symmetry. To address this, we construct a PT-symmetric operator $\tilde{U}(k_x, k_y, t)$ by smoothly deforming $U(t)$. This operator $\tilde{U}(k_x, k_y, t)$ preserves the phase band structure (including its singularities) and satisfies $\tilde{U}(k_x, k_y, t) = U(k_x, k_y, t)$ at $t = 0$ and $t = T$. Formally, $\tilde{U}(k_x, k_y, t)$ can be expressed through spectral decomposition as:

$$\tilde{U}(k_x, k_y, t) = \sum_{n=1}^{3} e^{-i\phi_n(k_x, k_y, t)} |\psi_n(k_x, k_y, t)\rangle\langle\psi_n(k_x, k_y, t)| \tag{13}$$

where $e^{-i\phi_n(k_x, k_y, t)}$ denotes the eigenvalues of $\tilde{U}(k_x, k_y, t)$ and $\phi_n(k_m, t) \in (-\pi/T, \pi/T]$ ($m = x, y$) forms the phase bands in the three-dimensional (3D) momentum time space.

As the time progresses, the phase bands may undergo overlapping and subsequent separation, leaving behind degenerate Dirac points within the band gaps in the three-dimensional momentum-time space. At time $t = T$, the phase bands transform into quasienergy bands. In the anomalous phase, the phase bands within each band gap undergo fourfold crossings during the course of time evolution. Here, we illustrate the anomalous phase at $t = T$, as depicted in Fig. 4(a). For ease of reference, we label the energy bands from bottom to top as the first, second, and third bands, while keeping in mind their periodic replication properties. The band gaps between the first and second bands, the second and third bands, and the third and first bands (accounting for periodicity) are subsequently denoted as the first, second, and third band gaps, respectively.

We observe that the phase bands undergo fourfold contacts in all three band gaps, resulting in the formation of three sets of Dirac points. In the Fig. 4(a), we have marked the four Dirac points in the first band gap with yellow circles. The Dirac points in the second and third band gaps are labeled with blue triangles and red pentagrams, respectively. The presence of Dirac singularities in the phase bands induces the emergence of edge states within the corresponding band gaps. Given the intricate interweaving of these three band gaps, this also elucidates why the conventional Chern number cannot be employed for calculating the topological invariants of the two-dimensional non-Abelian Floquet structures; instead, a composite Chern number should be utilized.

The Dirac points between energy bands in Fig. 4(a) correspond one-to-one with the magnified fourfold degeneracy observed in the quasienergy spectrum of Fig. 4(b). Fourfold degeneracy implies that four eigenstates share identical quasienergy values under specific conditions, indicating quantum state degeneracy. The Dirac points in the first and second band gaps respectively correspond to four degenerate eigenstates with quasienergies $E/\pi = -0.102$ and $E/\pi = 0.102$ in the quasienergy spectrum. The Dirac points in the third band gap correspond to four adjacent degenerate eigenstates with quasienergies $E = \pi$ and $E = -\pi$ in the quasienergy spectrum.

The presence of Dirac singularities in the phase bands induces the emergence of edge states within the corresponding band gaps. In fact, these edge states correspond to the degenerate eigenstates observed in the quasienergy spectrum of Fig. 4(b). In Fig. 4(c-e), we investigate the spatial distribution of eigenstates to visualize the boundary states. These boundary states correspond to quasienergies $E/\pi = -0.102$, $E/\pi = 0.102$, and $E/\pi = \pm 1$. For each quasienergy, there are four distinct special eigenstates (i.e., boundary states). We present only one representative eigenstate diagram for each quasienergy, while the remaining nine eigenstate diagrams are provided in Supplementary Material Appendix B. Therefore, the formation and evolution process of Dirac points (as a function of time) clearly elucidate the dynamical origin of edge states and determine their emergence in different band gaps. While the composite Chern number can only indicate the presence or absence of edge states via "non-zero/zero" classification, it fails to distinguish their specific band gap locations or quantify their multiplicity.

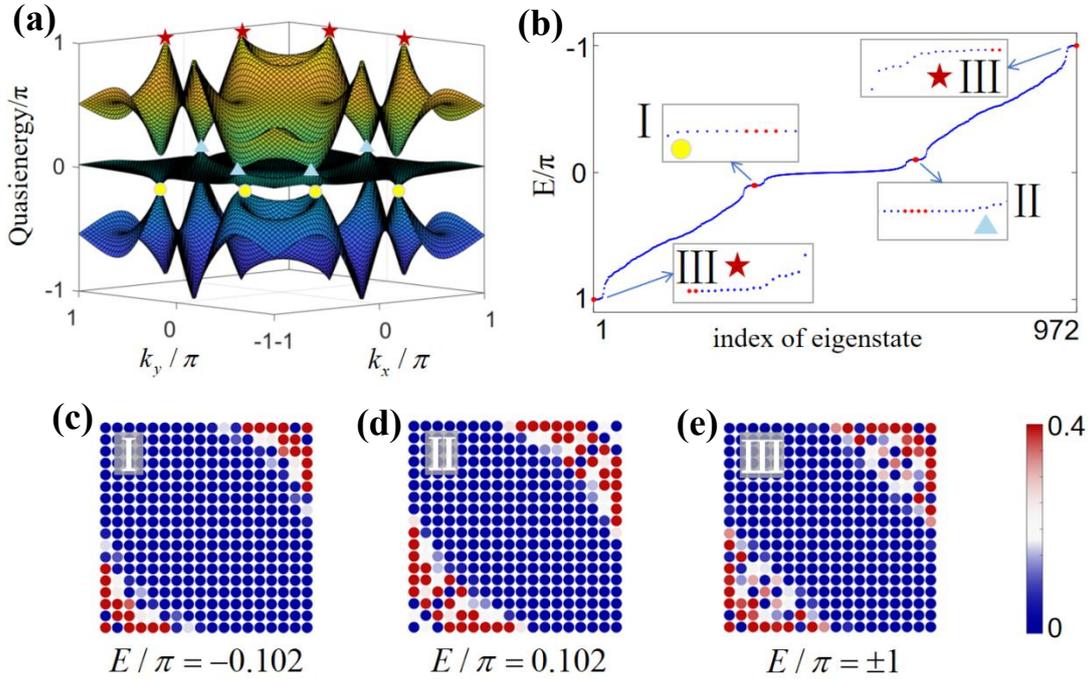

**FIG. 4.** FNATI with anomalous edge states. (a) Phase bands of $\tilde{U}(k_x,k_y,t)$ in the 3D momentum-time space. The Dirac-point singularities are marked with stars, triangles, and circles. (b) Zoomed-in Floquet Spectrum in the Quasienergy Zone Space. The Floquet Spectrum displays the presence of four-fold degeneracy in the topological modes. Separate occurrences are observed for the three types of four-fold degeneracy. (c-e) Spatial distribution of special eigenstates in Floquet systems. The boundary states in these three figures correspond to the **I**, **II**, and **III** types of four-fold degeneracy scenarios, respectively.

# IV CONCLUSION

We conducted a systematic investigation into the topological properties of two-dimensional FNATIs with trivial quaternion charges. Our work reveals differences in the BBC between static and Floquet-driven systems. The spatial exchange interaction of the driving field can excite stable interface states. Moreover, Floquet driving can induce degenerate Dirac points in the three-dimensional momentum-time space. Additionally, we confirmed that the composite Chern number can effectively characterize the overall topological non-triviality of two-dimensional three-band systems with degenerate bands. These findings simultaneously underscore the complementary roles of topological invariants and dynamical singularities in characterizing non-Abelian topological phases.

Our work extends the research on FNATIs to two-dimensional systems, establishing an intrinsic connection among time-driven dynamics, spatial exchange interactions, and topological states, thereby providing a theoretical framework for high-dimensional non-equilibrium topological phases. Subsequent research can further delve into the unique properties and universal laws of high-dimensional Floquet non-Abelian topological phases, as well as explore more novel topological states and their potential applications. Meanwhile, it is hoped that this work will inspire experimental investigations on platforms such as photonic lattices [71-73] and ultracold atoms [74-78]. Through precise manipulation of experimental parameters, the preparation and observation of FNATIs can be achieved, facilitating the transition of topological quantum information and waveguide technologies [79-82] from theory to practical applications and injecting new vitality into the development of quantum technologies [83, 84] in the future.

# ACKNOWLEDGMENTS

The authors thank for the support by National Natural Science Foundation of China under (Grant 12404365).

# APPENDIX A: BULK-BOUNDARY CORRESPONDENCE UNDER DIFFERENT DRIVING SCHEMES

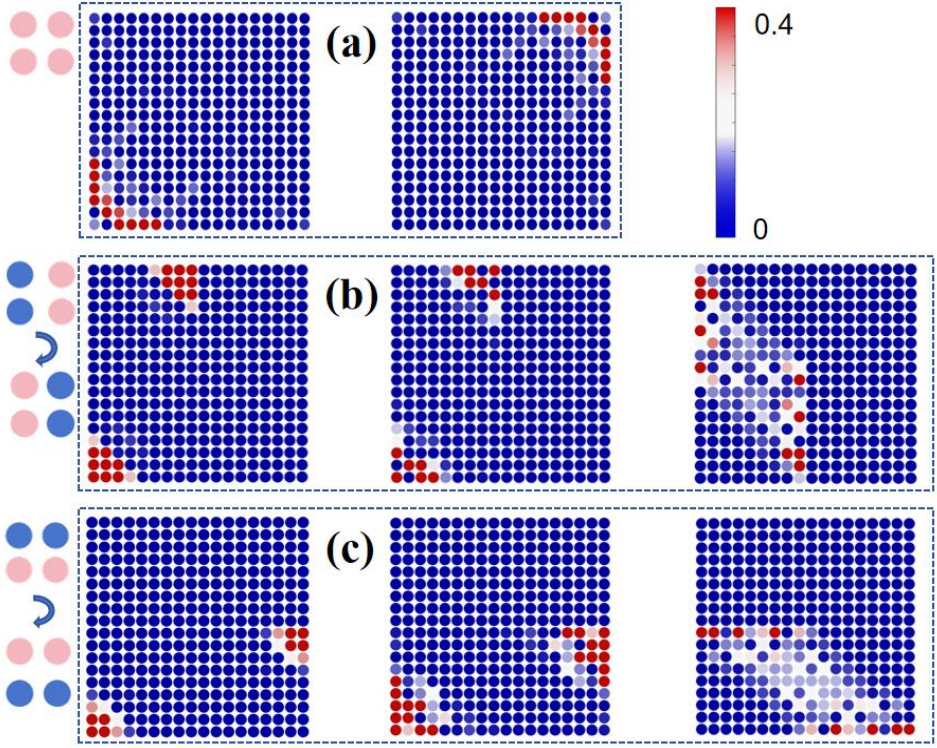

**FIG. S1.** BBC and interface modes induced by different driving modes. (a) Spatial distribution of special eigenstates in static systems. The left and right panels illustrate two types of edge states emerging during the evolution process. (b-c) Spatial distribution of special eigenstates in Floquet systems under spatial exchange driving. From left to right, they are the corner states, edge states, and interface modes emerging during the evolution, respectively.

Beginning with Fig. S1(a), we observe the emergence of boundary states in a static system governed by the Hamiltonian. Specifically, in the two scenarios depicted in the figure, energy accumulates at edge-2 and edge-1, respectively. Turning to Fig. S1(b), we illustrate a situation where, over the course of a complete period, the Hamiltonian evolution paths on the left and right sides of the system follow path $H_1 \to H_2 \to H_1$ and path $H_2 \to H_1 \to H_2$, respectively. Consequently, energy accumulates at the corner states, edge states, and interface modes on the left side. Finally, in Fig. S1(c), we observe corner states, boundary states, and interface modes. This is attributed to the effect of a spatially upper-lower-swapped driving mode, with energy concentrated on the lower side.

# APPENDIX B: SPECIAL EIGENSTATE CORRESPONDING TO FOURFOLD DEGENERACY

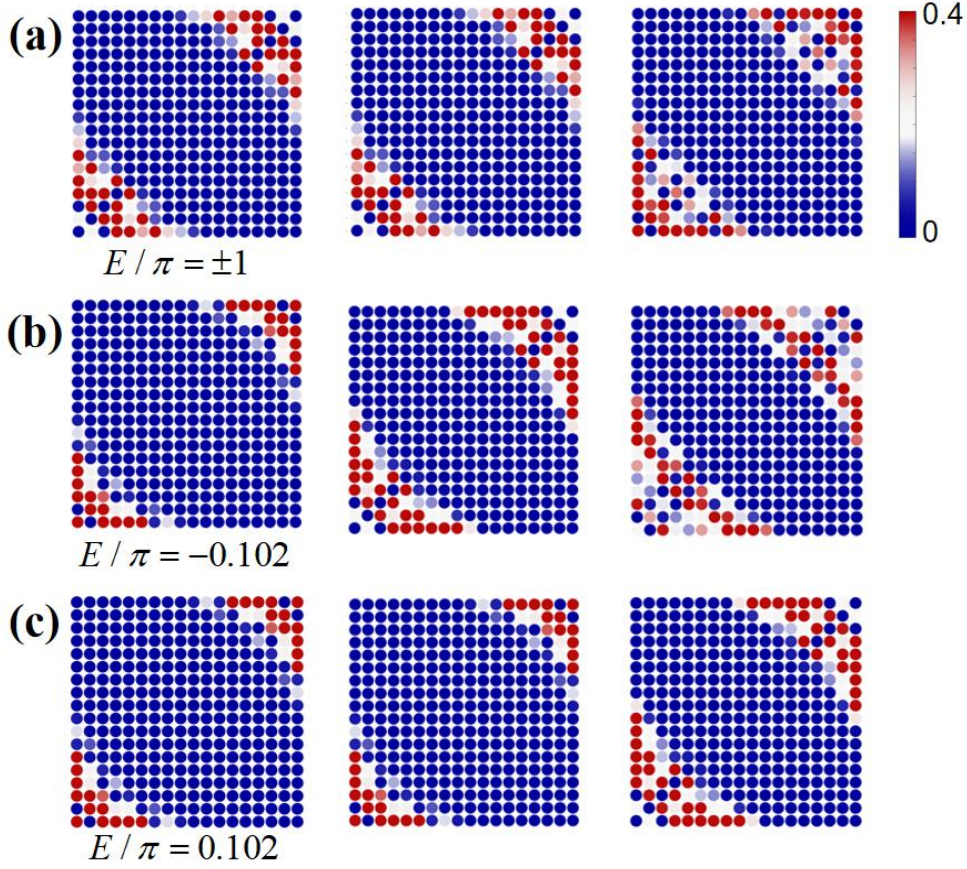

**FIG. S2.** Spatial distribution of special eigenstates in Floquet systems. (a-c) Eigenstates corresponding to quasienergies $E/\pi = \pm 1$, $E/\pi = -0.102$, and $E/\pi = 0.102$.

The special eigenstates are determined by phase-band singularities. In the main text, we found that fourfold degeneracy occurs in each energy gap, meaning four Dirac points (singularities) emerge within every gap. Consequently, each quasi-energy corresponding to a Dirac point should be associated with four distinct special eigenstates (corner or edge states). The energy distributions in FIG. S2(a-c) provide supplementary information on the special eigenstates corresponding to the quasi-energies $E/\pi = \pm 1$, $E/\pi = -0.102$, and $E/\pi = 0.102$, as discussed in the main text.

# APPENDIX C: TEMPORAL TRAJECTORY OF ENERGY DISTRIBUTION

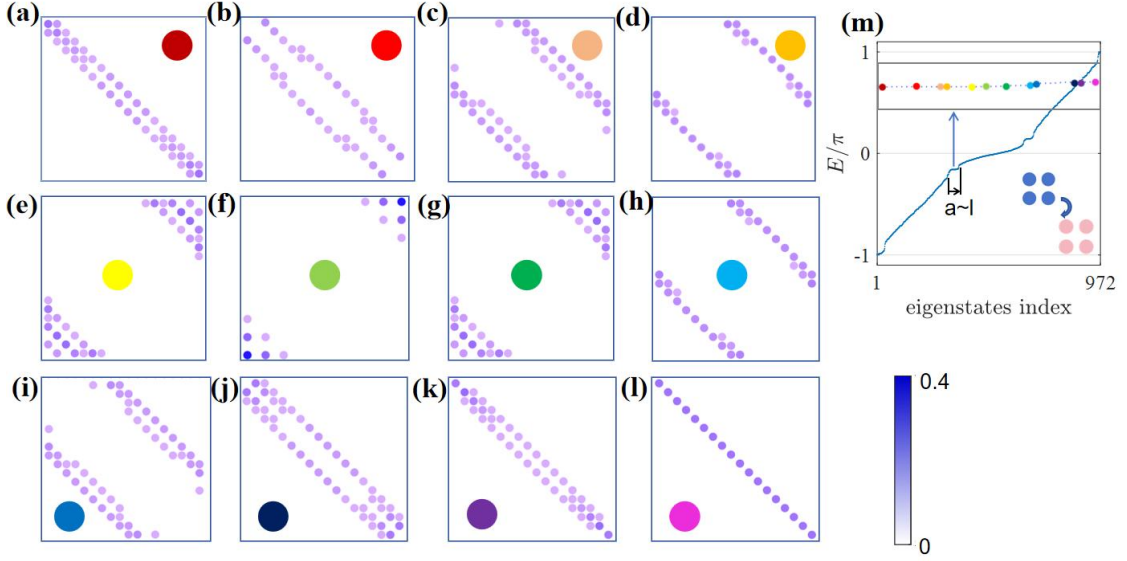

**FIG. S3.** Temporal evolution of energy distribution profiles. (a-l) The eigenstate distributions corresponding to different energy values. (m) Zoomed-in Floquet Spectrum in the Quasienergy Zone Space. The arrow directions in the figure indicate the sequential order in which the quasi-energies, corresponding to panels (a) through (m), appear in the quasienergy spectrum.

In a Floquet system driven by the alternating application of Hamiltonians $H_1$ and $H_2$, we observe the emergence of corner states at quasienergies $E/\pi = -0.102$, $E/\pi = 0.102$, and $E/\pi = \pm 1$. Furthermore, during the evolution of eigenvalues, certain regular patterns in the energy distribution are identified. We observe that, in FIG. S3(a-l), in the vicinity of energies corresponding to the corner (or edge) states, the energy distribution consistently demonstrates a tendency: either to accumulate from the corner states towards the diagonal or from the diagonal towards the corner states. These energy distribution plots successively correspond to the eigenstates emerging along the arrow direction in FIG. S3(m), representing the energy variations during the evolution process near the energy level $E/\pi = -0.102$.